# Mixed polytype/polymorph formation and its effects on the electronic properties in InSe films grown by molecular beam epitaxy on GaAs(111)B

Maria Hilse,[1,2,*] Justin Rodriguez,[2,3] Jennifer Gray,[2] Jinyuan Yao,[2,3] Shaoqing Ding,[2,3] Derrick Shao Heng Liu,[1,2] Mo Li,[4] Joshua Young,[4] Ying Liu[2,3,×], and Roman Engel-Herbert [1,2,5,†]

[1] Department of Materials Science and Engineering, The Pennsylvania State University, University Park, PA 16802, USA.

[2] The Materials Research Institute, The Pennsylvania State University, University Park, PA 16802, USA.

[3] Department of Physics, The Pennsylvania State University, University Park, PA 16802, USA.

[4] Otto H. York Department of Chemical and Materials Engineering, New Jersey Institute of Technology, Newark NJ, 07102, USA.

[5] Paul-Drude-Institut für Festkörperelektronik, Leibniz-Institut im Forschungsverbund Berlin e.V., Berlin, Germany.

* mxh752@psu.edu

* yxl15@psu.edu

† engel-herbert@pdi-berlin.de

**Abstract**

The top-down synthesis of inherently ferroelectric semiconductors and their integration with traditional material platforms have the potential to enable new low power logic devices, and to harness the bulk photoelectric effect for more efficient photovoltaic cells. InSe is a layered van der Waals compound exhibiting multiple polytypes, with semiconducting γ-InSe revealing a non-centrosymmetric space group and showing a high carrier mobility at room temperature. Here we report the growth of InSe films on close to lattice matched semi-insulating GaAs(111)B substrates by molecular beam epitaxy (MBE). Excellent nucleation behavior resulted in the growth of smooth, single phase InSe films. The dominant polytype determined from X-ray diffraction was the targeted γ-InSe, however Raman spectroscopy revealed spatial variations in the overall low-intensity non-centrosymmetric vibration modes. Transmission electron microscopy uncovered the presence of the three bulk polytypes β, γ, and ε-InSe coexisting in the films arranging in nanosized domains. The different polytypes can be interpreted as sequences of stacking faults and rotational twin boundaries of γ-InSe made from individual non-centrosymmetric Se-In-In-Se layers with $P\bar{6}m2$ symmetry. A second, centrosymmetric Se-In-In-Se layer polymorph was identified with $P\bar{3}m$ symmetry, which is typically not present in InSe bulk phases. First principles calculations revealed small formation energy differences between the InSe polymorphs and polytypes, yet sizeable differences in their electronic properties. Nanoscale domain sizes of varying polytypes thus resulted in sizeable electronic disorder in the grown films that dominated the electronic transport properties. Our results indicate that bottom-up thin film synthesis is a viable synthesis route towards stabilization of InSe polytypes not present in the bulk. An improved understanding and control over InSe growth conditions is, however, required to stabilize the polymorph of choice, and to ultimately inscribe a specific layer sequence on demand by utilizing the bottom-up layer-by-layer growth mode capability available in MBE.

## Introduction

The discovery of graphene[1] brought dramatic advances not only in the fundamental studies of two-dimensional (2D) materials, but also in developing novel device concepts.[2–8] Monolayer graphene obtained by mechanical exfoliation of a single bulk crystal has been shown to possess a carrier mobility exceeding $10^4$ cm$^2$V$^{-1}$s$^{-1}$ at room temperature, owed to the linear electron dispersion at the Fermi level. Other 2D materials with a finite energy gap such as black phosphorus[9] were also found to possess a high carrier mobility and transition metal chalcogenides[10–16] have revealed promising optoelectronic properties while maintaining a sizeable carrier mobility. The weak interlayer bonding common to these 2D material systems makes them highly desirable for applications, as it allows to combine different functionalities at the nanoscale with tunable coupling between the individual layers while easing the materials integration challenge. Ultimately, a reliable growth of these materials that retains their favorable transport and optoelectronic properties in wafer-size films is highly desirable for their technological applications.[17]

To that end, the layered compound InSe has attracted much attention in recent years due to the reported high carrier mobility and promising optoelectronic properties.[18–25] Among the bulk-stable polytypes $\beta$-, $\gamma$-, and $\varepsilon$-InSe, $\gamma$-InSe possesses a direct energy band gap of about 1.26 eV in the bulk, which increases with reducing the number of InSe layers and finally reaches 2.11 eV in the single quadruple (Se-In-In-Se) layer limit. The energy gap is indirect for single layer (SL) InSe.[23,24,26] Carrier mobilities higher than 1000 cm$^2$V$^{-1}$s$^{-1}$ at room temperatures have been reported,[19,27–30] which were found to depend on the substrate[28,31,32] and changed when the film was encapsulated with another 2D crystal.[33–35] The mobility reported in thin films grown on various substrates tended to be lower than values found using mechanically exfoliated crystals.[24,29,30] This indicates that either the defects introduced by mechanical exfoliation are potentially not detrimental to the carrier transport characteristics, or the bottom-up growth of InSe gives rise to a high defect concentration that is currently not well understood. The closely related layered van der Waals (vdW) chalcogenide $In_2Se_3$ composed of quintuple layers [Se-In-Se-In-Se] was furthermore reported to be ferroelectric in at least two of its polytypes.[36–43] Since two of the three bulk-stable InSe polytypes γ and ε are non-centrosymmetric[20,44,45] and thin crystals of β-InSe were reported to be ferroelectric when strained,[45,46] it poses the question whether ferroelectricity can be present in InSe as well.

Here, we report the growth of InSe films by molecular beam epitaxy (MBE) on close to lattice matched semi-insulating GaAs(111)B. Native oxide removal of GaAs(111)B surface was optimized along with film growth parameters using reflection high energy electron diffraction (RHEED), X-ray diffraction (XRD), and atomic force microscopy (AFM). It is shown that the dominant InSe polytype in the films was γ-InSe. Polytype domain sizes were found to have nanoscale dimensions. In addition to different polytypes, the growth conditions far away from equilibrium gave rise to the abundant formation of a different InSe polymorph as well, where the tri-fold In-Se bonds in the upper Se monolayer was rotated by 180° relative to the tri-fold In-Se bonds in the lower Se monolayer, collapsing the non-centrosymmetric $P\bar{6}m2$ into the centrosymmetric $P\bar{3}m$ space group. Nanoscale polytype and polymorph domain arrangements were identified to cause electronic disorder, and implications of such on the electronic properties are discussed. Our results show that indeed bottom-up thin film synthesis is a viable synthesis route towards stabilization of InSe polytypes and polymorphs that are not present in the bulk. An improved understanding and control over InSe growth conditions is necessary to stabilize the [Se-In-In-Se] layer polymorph of choice and to suppress the nanoscale polytype domain formation. The ability to stabilize centrosymmetric quadruple layer and inscribing a specific layer sequence on demand utilizing the layer-by-layer growth mode available in MBE is a suitable route towards engineering functional stacks of InSe polytypes offering superior electronic properties over their centrosymmetric counterparts.

**Results and discussion**

Polytypism in InSe

Fig. 1 shows the atomic structure of the bulk stable InSe polytypes referred to as β, γ, and ε-InSe.[20,25,31,47–49] The sp$^3$ hybridization present in InSe is indicative of the strong covalent bond character within each InSe layer consisting of the four individual atomic layers [Se-In-In-Se]. While the outer ones are made exclusively from Se, the two inner ones only consist of In. Direct In-In bonds in the center as well as Se-In bonds tie these four atomic layers together, while only weak vdW interactions are present between adjacent Se layers. The In-In bond direction is normal to the quadruple layer, and commonly the trifold In-Se rotational symmetry of the sp$^3$ bond geometry remains the same on either end of the bonded In-In atoms. This yields the $P\bar{6}m2$ space group for a single InSe layer. Note, that for the projection along the a-axis $[11\bar{2}0]$ one In-Se bond appears longer, while the opposite one appears shorter. The former has its In-Se bond direction perpendicular to the projection vector, while the latter has a non-zero projection onto the a-axis. All atoms of SLs projected along $[11\bar{2}0]$ thus form a chain of irregular hexagons much like a line cut from a slightly distorted honeycomb, which is highlighted by overlaying the SL atomic arrangement in the [11$\bar{2}$0] zone axis of Fig. 1 with irregular hexagons.

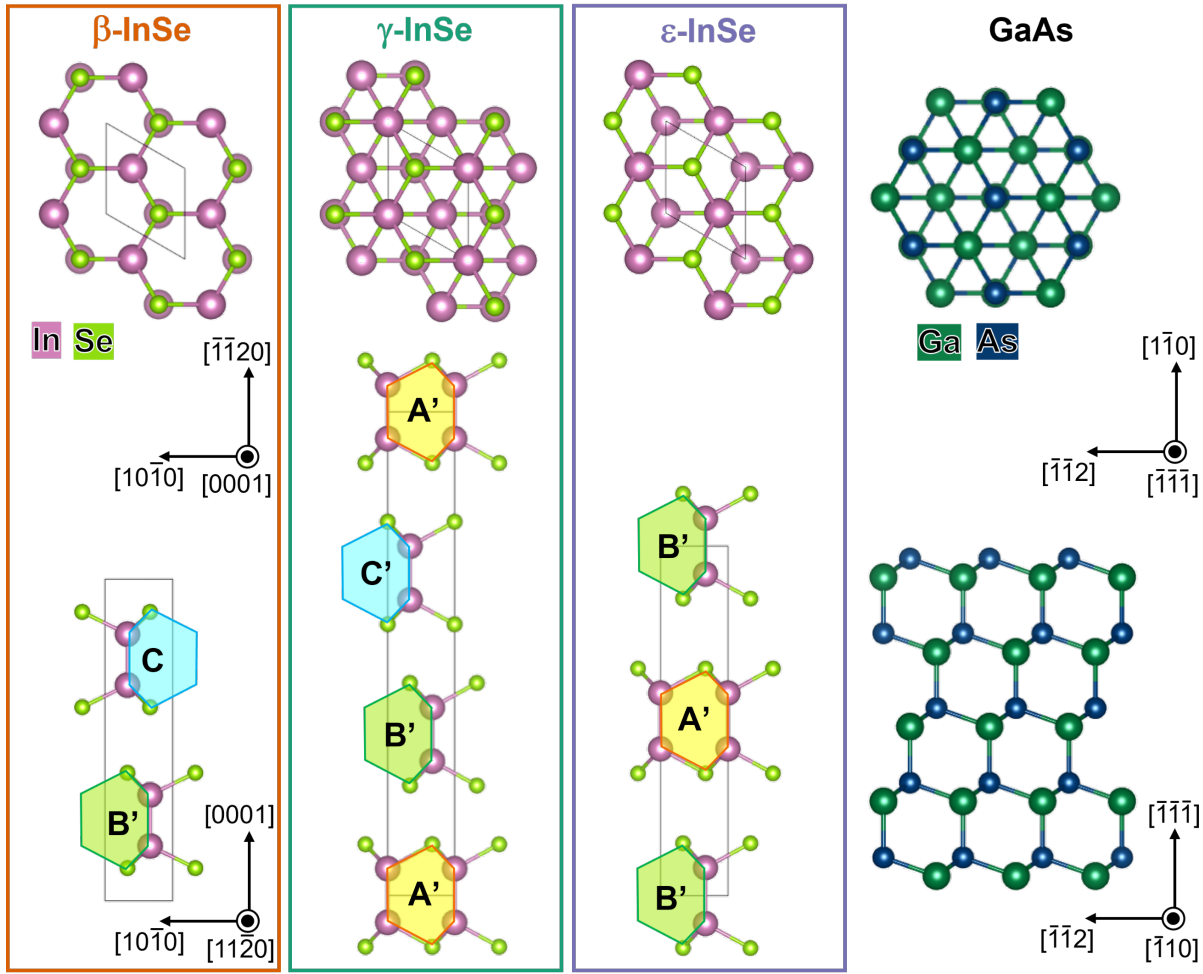

**Figure 1: Crystal structure of InSe polytypes and the GaAs substrate.** *Projection of the InSe crystal structures for the different bulk-stable polytypes β, γ, and ε were chosen along [0001] (top row) and [11$\bar{2}$0] (bottom row). The bonding configuration within each InSe single layer (SL) is highlighted by irregular hexagons. Note, how the two shorter In-Se bonds in this projection provide the appearance of a C. Stacking sequences of InSe SL resulting from the relative in-plane alignment obtained by relative displacement of the neighboring SL by 1/3 of the in-plane primitive lattice vector labeled A, B, and C. A rotation of an InSe SL by 60° about the layer normal [0001] results in a 'flip' of the Se-In-In-Se C-shaped appearance, indicated by a prime. The GaAs substrate is illustrated along [$\bar{1}\bar{1}\bar{1}$] (top row) and [$\bar{1}$10] (bottom row).*

Owed to the weak vdW interaction between the quadruple [Se-In-In-Se] InSe SLs they stack in different sequences with different in-plane displacement relative to adjacent layers forming various polytypes. The InSe polytypes can be categorized by the relative translational and rotational relationship of adjacent InSe layers, i.e. the relative shift and rotation of the irregular hexagon chains. In the case of β-InSe neighboring layers along the stacking sequence are translated by 1/3 along the primitive lattice vector within the layer (position of lower SL irregular hexagon B changes to C position in the SL above) and rotated by 60° relative to one another (flipping of the irregular hexagon chain denoted by a prime in Fig. 1). This way the ridges and grooves of adjacent Se layers (irregular hexagons in Fig. 1) and both In and Se sites between neighboring SLs are aligned. This B'-C layer stacking sequence results in the centrosymmetric space group $2H - P6_3/mmc$ (No. 194) for β-InSe.[50–52] In contrast, γ-InSe has a longer stacking sequence. Here, the upper and lower neighboring layers of any given SL are not identical. Instead of combining a translation with a rotation, adjacent layers are only translated within the plane, the upper one by 1/3 along the primitive lattice vector (irregular hexagon position shifts from B to C in Fig. 1) and the lower one by -1/3 along the same primitive lattice vector (irregular hexagon position shifts from B to A in Fig. 1), yielding an A'-B'-C' (equivalent to A-B-C) stacking of the non-centrosymmetric space group $3R - R3m$ (No. 160).[50–52] In contrast to β-InSe, the missing rotation between SLs in γ-InSe only allows to have either the In position lined up with the Se position of one neighboring SL, or the Se position line up with the In position of the other neighboring SL, but not both. Finally, for ε-InSe a bi-layer sequence A'-B' (or A-B) is formed by translating both, the upper and lower adjacent SL by 1/3 along the in-plane primitive lattice vector in the same direction (irregular hexagon position shift from A to B both going up and down the stacking sequence, i.e., along [0001] and [000$\bar{1}$], respectively in Fig. 1), resulting in the non-centrosymmetric space group $2H - P\bar{6}m2$ (No. 187).

[50–52] For this polytype the In positions line up with the Se positions going from B to A, and the Se position line up with the In position going from A to B. The in-plane lattice parameters are identical for all three polytypes (a=4.01 Å),[50–52] while the lattice parameters along the stacking sequence are multiples of the quadruple [Se-In-In-Se] layer and the van der Waals gap between them. For β-InSe and ε-InSe a nearly identical out-of-plane lattice parameter of c≈16.64 Å, and c≈16.70 Å was found, respectively,[50–52] while the A-B-C layer sequence of γ-InSe gave c≈24.95 Å.[50–52] The in-plane lattice parameter of all InSe polytypes is furthermore almost ideally matched to the in-plane lattice parameter of the GaAs(111) plane. GaAs has a lattice parameter of a=5.653 Å in the zincblende structure resulting in atomic spacings of 3.997 Å in the (111) plane. The three-fold symmetry of the (111) plane along with a small lattice mismatch between GaAs(111) and InSe of about -0.3% impose a compressive in-plane strain on InSe.

InSe film growth and structural characterization

Fig. 2(a) shows the RHEED images taken before and after the native oxide removal of GaAs(111)B and after the growth of InSe films. Before oxide removal, the RHEED images were blurry with a high background intensity. Faint and diffusive Kikuchi lines, and diffraction rods of low intensity along both azimuths presented initially with no discernable surface reconstruction evidencing an amorphous native oxide overlayer. RHEED image quality dramatically improved after exposing the GaAs(111)B surface to a flux of highly reactive hydrogen. Highly intense and sharply defined diffraction rods of a (1×1) reconstructed GaAs(111)B surface were found, including sharp Kikuchi lines and diffraction features of the second Laue circle. The native oxide overlayer thickness was dramatically reduced and a high degree of crystalline order of the

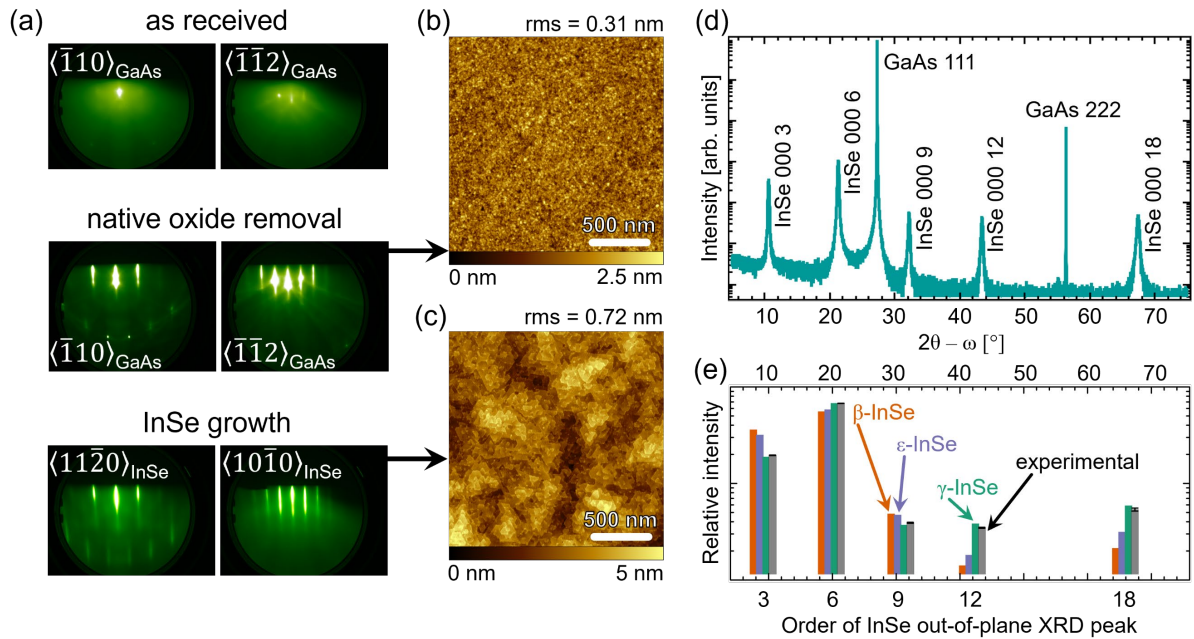

**Figure 2: Structural characterization.** *(a) RHEED images along the high-symmetry directions of as received GaAs(111)B substrates, the substrate after the native oxide removal in the MBE, and the InSe film after growth. (b) Surface topography as it presented in AFM of the GaAs(111)B substrate after native oxide removal in the MBE. (c) AFM image of the InSe film morphology. (d) On-axis XRD scan of the InSe thin film grown on GaAs(111)B. (e) Calculated relative structure form factors of β-, γ-, and ε-InSe and measured relative XRD peak intensities of the grown InSe film extracted from (d).*

GaAs(111)B was achieved. The surface morphology of the GaAs(111)B substrate right after the native oxide removal was measured by AFM within 20 min of the wafer being taken out of the MBE chamber. As shown in Fig. 2(b), a smooth substrate surface with a grainy texture was observed. The root means square (rms) surface roughness was 0.31 nm. Finally, the observed RHEED images after InSe growth shown in Fig. 2(a) were taken along the same high symmetry azimuths $\langle\bar{1}10\rangle$ and $\langle\bar{1}\bar{1}2\rangle$ of the GaAs(111)B. Sharply defined diffraction rods of high intensity in front of a low intensity background indicated single crystal InSe growth. The low surface energy of the vdW gap naturally resulted in the growth of InSe along the c-axis, i.e. GaAs($\bar{1}\bar{1}\bar{1}$) || InSe(0001). The in-plane epitaxial relationship observed in RHEED was GaAs[$\bar{1}$10] ||

InSe[11$\bar{2}$0]. It should be noted that the width of the InSe diffraction rod was narrower than the GaAs diffraction rod, indicating a larger lateral coherency of the InSe surfaces.

The InSe film surface morphology determined by AFM is shown in Fig. 2(c). The rms roughness was increased more than twofold (0.72 nm). Rather than a grainy texture atomically smooth triangular shaped islands were observed. About eight different SL InSe layer levels were found within the 2μm × 2μm AFM scan, indicating a limited lateral diffusivity during growth. It is anticipated that higher growth temperatures would allow for larger InSe layer terrace widths.

On-axis XRD scans are shown in Fig. 2(d). The total InSe film thickness was found to be about 40 nm, determined from the full width half maximum XRD 2θ film peaks, yielding a growth rate of about 0.21 Å/s. Aside from the GaAs(111) substrate peak all XRD film peaks could be assigned to reflection arising from InSe set of basal planes 000 m stacked along the c-axis. In case of the γ-InSe polytype with a three-layer stacking sequence m=3×n ($n \in N$) the 000 15 reflection is being suppressed limiting the observed XRD peaks to orders m=3,6,9,12, and 18 in the probed 2θ range of Fig. 2(d) as indicated,[50–52] while for the two bi-layer stacking sequence polytypes β-InSe and ε-InSe m=2×n ($n \in N$) the 000 10 reflection is being suppressed limiting the observed XRD peaks to orders m=2,4,6,8, and 12 in the probed 2θ range of Fig. 2(d) if β-/ε-InSe were used for peak labeling.[53–55]

The degree of polytypism present in the grown films was approximated by comparing the normalized structure form factors (each structure form factor normalized by the sum over all structure form factors for the different reflection orders considered for each polytype) of the X-ray reflections. Fig. 2(e) shows the experimentally determined (gray) and for the different polytypes (orange - β-InSe, purple - ε-InSe, and green - γ-InSe) calculated relative intensities, i.e., normalized structure form factors $i_m = I_m / \sum_k I_k$ for the different reflection orders $m$ plotted on a logarithmic scale versus the diffraction angle 2θ (top x-axis) and the diffraction peak order $m$ (bottom x-axis assuming γ-InSe nomenclature). The absolute values of relative XRD intensities (rel. XRD intensities), and normalized structure form factors (rel. SF) for each polytype as well as the respective XRD peak and peak position in 2θ are summarized in Table 1. All experimental values were obtained by fitting a Voigt profile to the individual X-ray film diffraction peaks after background subtraction and normalizing the integrated peak intensity from the fit of each peak to the sum of all InSe peak intensities.

The closest match between the experimentally determined relative X-ray intensities and structure form factors was found for γ-InSe. While the experimentally determined relative X-ray intensities were slightly larger compared to the normalized structure form factor of the γ polytype for the 000 3 and 000 9 reflections, the intensities were somewhat smaller for the 000 12 and 000 18 reflections. This reflected the trend of both bi-layer stacked polytypes β- and ε-

**Table 1.** *Experimentally determined XRD peaks and peak positions 2θ, normalized structure form factors (rel. SF) of β-, γ-, and ε-InSe polytypes, experimentally determined relative X-ray intensities (rel. XRD intensities), and calculated polytype fraction of the β-polytype in γ-InSe assuming exclusively γ- and β-InSe presence in the film (polytype γ-β), and the fraction the ε-polytype in γ-InSe assuming exclusively γ- and ε-InSe (polytype γ-ε).*

| XRD peak and position 2θ [°] | rel. SF β-InSe | rel. SF γ-InSe | rel. SF ε-InSe | rel. XRD intensities | polytype γ-β [%] | polytype γ-ε [%] |
|---|---|---|---|---|---|---|
| 000 3 - 10.602 | 0.3613 | 0.1881 | 0.3193 | 0.196 ± 0.001 | 4.5 ± 0.9 | 6 ± 1 |
| 000 6 - 21.313 | 0.5548 | 0.6774 | 0.5840 | 0.677 ± 0.002 | 1 ± 1 | 1 ± 2 |
| 000 9 - 32.200 | 0.0485 | 0.0371 | 0.0472 | 0.0391 ± 0.0005 | 18 ± 5 | 20 ± 5 |
| 000 12 - 43.416 | 0.0141 | 0.0384 | 0.0181 | 0.0348 ± 0.0004 | 15 ± 2 | 18 ± 2 |
| 000 18 - 67.410 | 0.0213 | 0.0589 | 0.0314 | 0.054 ± 0.002 | 12 ± 5 | 17 ± 6 |

InSe, which compared to γ-InSe have structure form factors larger for the 000 3 and 000 9 reflections, and smaller for 000 6, 000 12 and 000 18 reflections. From the X-ray analysis it was concluded that γ-InSe was the dominant polytype for the growth conditions, but that films did not exclusively contain this single polytype. It was noted that β-InSe had a larger normalized structure form factor than ε-InSe for the 000 3 and 000 9 reflections, and conversely a smaller normalized structure form factor for the 000 6, 000 12, and 000 18 reflections, respectively. Since the normalized structure form factor of ε-InSe was closer to γ-InSe the approximation of only γ- and ε-InSe polytypes contained in the film will provide an upper bound estimate for the degree of polytypism. Conversely, a conservative estimate – a lower bound estimate for the degree of polytypism is expected if the film is assumed to only contain γ- and β-InSe polytypes. The upper (lower) bound of mixed polytypism $d_{\gamma-\varepsilon}$ ($d_{\gamma-\beta}$) in the film was obtained by expressing the relative X-ray reflection intensity $i_{exp}$ for a specific order of reflection as a sum of the normalized structure form factors of the γ-InSe and the ε-InSe (β-InSe) polytype: $i_{exp} = (1 - d_{\gamma-\varepsilon}) \cdot i_{\gamma} + d_{\gamma-\varepsilon} \cdot i_{\varepsilon}$, and $i_{exp} = (1 - d_{\gamma-\beta}) \cdot i_{\gamma} + d_{\gamma-\beta} \cdot i_{\beta}$, respectively. Table 1 summarizes the degree of polytypism determined from the relative X-ray intensities of all scattering angles. The spread in contained polytypism in the sample depended quite strongly on the order of diffraction, which was attributed to the inherent limitation of the method, the introduction of systematic error in the Voigt profile fitting and background subtraction, or the neglect of additional polytypes present in the film with different normalized structure form factors. In all cases, polytypism was experimentally confirmed by this analysis, which is most reliable for the higher order reflections where a large difference between the normalized structure form factors for the different polytypes was present.

Complementing structural confirmation of type and amount of InSe polytypism present in the films can be obtained from Raman measurements. More localized information due to the focused light spot provided first insights into polytype domain size and arrangement and whether a single polytype stacking can be obtained throughout the entire film thickness. Fig. 3(a) shows the typical Raman spectrum obtained from InSe films on GaAs(111)B in the spectral range between 90 cm$^{-1}$ and 310 cm$^{-1}$. The following Raman modes related to InSe were identified: $A_{1g}^1$ (115 cm$^{-1}$), $E_{2g}^1$ (177 cm$^{-1}$), $A_{2g}^1(LO)$ (201 cm$^{-1}$), $E(LO)$ (210 cm$^{-1}$), and $A_{1g}^2(LO)$ (226 cm$^{-1}$).[20,56,57] The strong modes at higher frequencies of 267 cm$^{-1}$ and 291 cm$^{-1}$ were from GaAs, namely the transversal optical (TO) and longitudinal optical (LO) mode, respectively.[58] While the InSe Raman modes $A_{1g}^1$, $E_{2g}^1$, and $A_{1g}^2(LO)$ are present in all three InSe polytypes, the $A_{2g}^1(LO)$ and $E(LO)$ vibration modes are only observed for the non-centrosymmetric phases γ-InSe and ε-InSe.[44,45,59] This suggests that indeed non-centrosymmetric polytypes were present in the MBE grown InSe film, a mandatory but not sufficient condition to establish ferroelectric functionality of the films. The spatial homogeneity of the non-centrosymmetric polytype presence in the InSe film was obtained by taking Raman spectroscopy scans from 185 cm$^{-1}$ to 215 cm$^{-1}$ at spots 2.5 µm apart on a Hall bar device used for transport measurements on InSe, see further below. Fig. 3(b) shows an image of the Hall bar device taken by an optical microscope with a grid superimposed to indicate the locations at which the Raman spectra were taken across a (55×17.5) µm² area. All spectra taken at the respective points shown in Fig. 3(b) are superimposed in Fig. 3(c) with their backgrounds subtracted. Even though the $A_{2g}^1(LO)$ can be seen in the spectra the $E(LO)$ mode was hardly visible. The area under the $A^1_{2g}(LO)$ mode was used to quantify the spatial homogeneity of the $A^1_{2g}(LO)$ mode representing the amount of non-centrosymmetric polytypes in the film shown as a color-coded intensity map in Fig. 3(d). The Raman intensity of the non-centrosymmetric modes was very small and comparable to the signal noise, see Fig. 3(c). No significant changes were found in the Raman spatial map, which was attributed to either a too little signal, or too insignificant differences in the polytype presence at the length scales accessible by Raman maps.

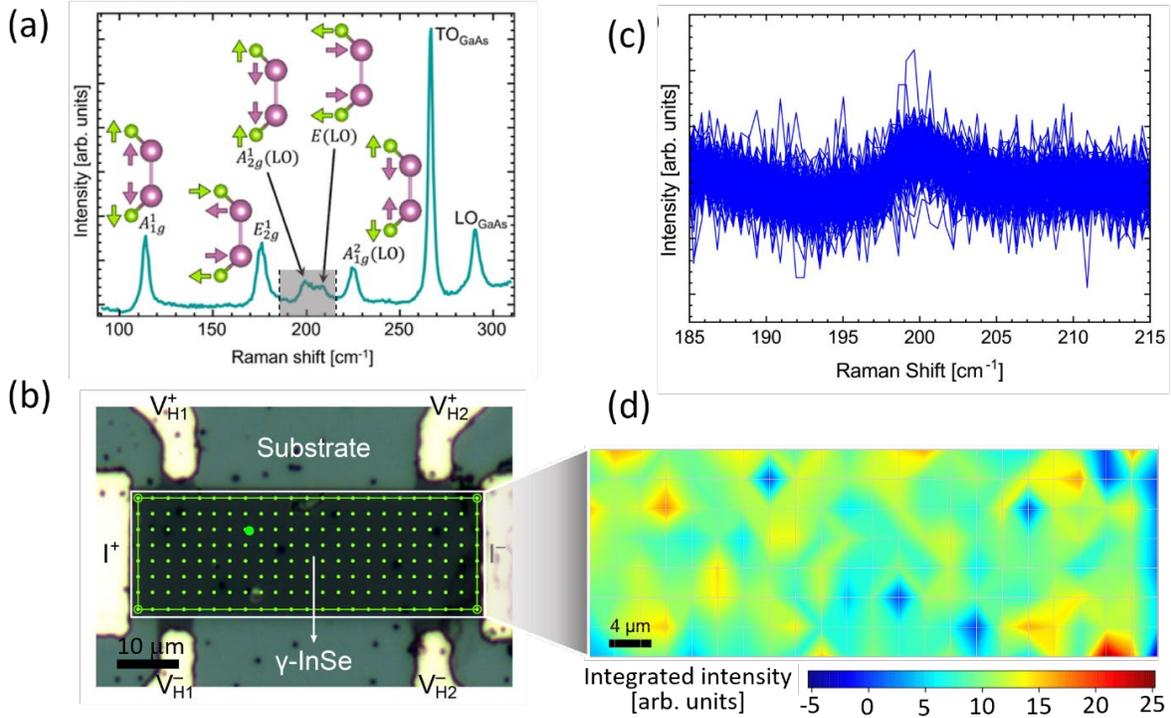

**Figure 3: Raman analysis.** *(a) Raman spectroscopy of an as-grown InSe film on GaAs(111)B. (b) Optical micrograph of the Hall bar device used for transport measurements displaying the electrode fingers in bright contrast contacting the rectangular InSe film channel overlayed with a 2.5µm-cell sized grid indicating the laser positions used to map the spatial homogeneity of the non-centrosymmetric polytype presence of InSe on GaAs(111)B between 185 cm$^{-1}$ and 215 cm$^{-1}$. (c) Raman spectra accumulated recorded at each grid point in (b). (d) Color-coded Raman intensity map of the vibrational mode at 185 cm$^{-1}$ plotted over the lateral dimensions of the grid in (b).*

## Microstructure of polytype/polymorph domains in InSe

Limited insights from XRD and Raman spectroscopy motivated to look further into the nanoscale arrangement of InSe polytypes in the film. A representative high-angle annular dark field (HAADF) high-resolution transmission scanning electron microscopy (STEM) image of an InSe film on GaAs is shown in Fig. 4(a). The higher Z number of In ($Z_{In}$=49) and Se ($Z_{Se}$=34) compared to Ga ($Z_{Ga}$=31) and As ($Z_{As}$=33) gave rise to a higher intensity of the film. The InSe revealed the layered structure arising from the highly anisotropic bond geometry. Twenty-one [Se-In-In-Se] quadruple layers were counted in the cross-section image in Fig. 4(a) separated by vdW gaps. The projection of In and Se rows shown in Fig. 4(a) corresponded to the $[11\bar{2}0]$ zone axis of InSe thus confirming the in-plane epitaxial relationship already found in RHEED. GaAs was identified from the parallelogram arrangement, where in the $[\bar{1}10]$ zone axis columns of Ga atoms cannot be distinguished from As ones. An atomically sharp structural interface was formed presenting a slight gap between film and substrate indicated by the interface labeled arrows. The atomic spacing between Ga(As) and In(Se) atoms in GaAs and InSe, respectively, near the interface yielded values of (3.7±0.1) Å for both film and substrate, and (3.6±0.1) Å for the In(Se) atomic distances in InSe 20 SLs away from the interface. This is in good agreement with the expected unstrained atomic spacings in the $[\bar{1}10]/[11\bar{2}0]$ projections that calculate to 3.462 Å in GaAs ($a \times \sqrt{3/8}$) and 3.47 Å in InSe ($a \times \sqrt{3}/2$), highlighting the good lattice match between film and substrate.

While the transition between the two dissimilar structures across the interface was structurally abrupt, a more gradual transient of the chemical distribution was found. Specifically, the two

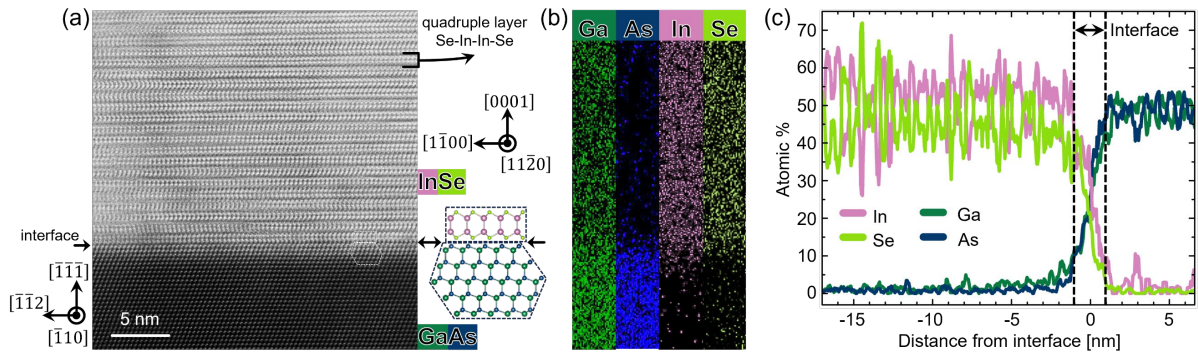

*Figure 4: Scanning transmission electron microscopy analysis. (a) High-resolution HAADF-STEM image of the InSe thin film grown on GaAs(111)B in cross section along the $[\bar{1}10]$ projection of GaAs. b) Elemental maps of Ga, As, In, and Se obtained by EDS from the STEM image in (a). (c) EDS line scans of the elemental distribution of Ga, As, In, and Se plotted over the distance from the interface of the STEM image in (a).*

Ga-As layers closest to the interface appeared brighter in HAADF-STEM, indicating that In was incorporated into the zincblende structure, but only protruded about two atomic layers deep into the substrate. This is expected from the grainy GaAs(111)B surface texture after native oxide removal using hydrogen and is similar to what was seen for GaSe growth on GaAs(111)B.[60] Elemental maps of Ga, As, In, and Se were recorded across the interface using energy-dispersive X-ray spectroscopy (EDS). Intensity maps of individual elements and cumulated line scans extracted from the maps are shown in Figs. 4(b) and 4(c) for each element. Elemental interdiffusion across the GaAs-InSe interface was near the detection limit of EDS in high-resolution STEM. Note the relatively high background of Ga in the InSe film, which was attributed to unintentional Ga deposition at the top and bottom surface of the cross-sectional specimen using a focused Ga ion beam for the STEM specimen preparation. As highlighted in Fig. 4(c), the interface region was less than 2 nm wide, a pronounced interdiffusion across the interface would present itself as a much wider interface region.

Close inspection of the structural relationship of adjacent InSe SLs in the film provided direct atomic scale insights into the polytype domain arrangement of the film. High-resolution STEM images taken from three different areas in the film are shown in Fig. 5. The relative in-plane shift of neighboring InSe SLs determined the stacking sequence and hence the InSe polytype present. The arrangement of In and Se in this projection were tracked in all presented images and compared to the respective polytypes discussed in Fig. 1. For ease of comparison the expected orientation and relative position of the irregular hexagons are highlighted in the legend of Fig. 5 and superimposed in the micrographs. Three lateral positions differing by an a/3 in-plane shift along the $[10\bar{1}0]$ direction gave rise to the ABC stacking sequence. Rotations of any of these layers about their normal axis by 60° 'flipped' the irregular hexagon, denoted by priming the capital letter labeling the respective SL, i.e., A → A'.

Starting from γ-InSe as dominant polytype with A-B-C-A stacking, ε-InSe can be derived from γ-InSe by incorporating a regular stacking fault removing either of the layers in an ordered fashion. For example, missing C layers give rise to the A-B-A stacking sequence. Equivalent sequences can be obtained by a ±a/3 in-plane shift, namely missing B or A layers yielding C-A-C or B-C-B stacking, respectively. Similarly, β-InSe can be derived from γ-InSe by simultaneously incorporating both, a regular stacking fault and a rotational domain boundary, whereby adjacent SLs were rotated by 60° in a regular fashion. For example, by removing the A and rotating the adjacent C layer the stacking sequence A-B-C-A of γ-InSe converts to the B-C'-B stacking of β-InSe. Therefore, a high density of planar defects within the film will give rise to many stacking faults and rotational domain boundaries within the stacking sequence yielding a nanoscale arrangement of different InSe polytypes. Specifically, considering area 1 shown in Fig. 5 (left column) the InSe stacking sequence from top to bottom can be interpreted as single polytype γ-InSe with a rotational domain boundary between SLs 1 and 2, followed by a stacking fault between SLs 2 and 3, a stacking fault with rotational domain boundary between SLs 6 and 7, followed by yet another rotational domain boundary between SLs 7 and 8, and a stacking fault to SL 9. An equivalent representation would be to assign layers 1 and 2 to the β-polytype, followed by γ-InSe (SLs 3-5 and SLs 9-15) that is interleaved with β-InSe (SLs 6-8). The stacking sequence found in area 2 was single polytype γ-InSe with a combined stacking fault and rotational domain boundary between SLs 8 and 9 and a stacking fault between SLs 9 and 10, as well as a rotational domain boundary between SLs 14 and 15. This can be alternatively assigned to a polytype sequence with γ-InSe (SLs 1-7), followed by β-InSe (SLs 8-9), ε-InSe (SLs 10-13), and back to β-InSe (SLs 14-15). The γ-InSe polymorph stacking sequence from area 3 top to bottom was interrupted by stacking faults between SLs 5 and 6, SLs 8 and 9, and SLs 11 and 12. A rotational twin boundary was found between SLs 10 and 11. The alternative and equivalent interpretation as nanoscale polytype arrangement yielded γ-InSe (SLs 1-4, and SLs 8-9), followed by ε-InSe (SLs 5-7), and β-InSe (SLs 10-11), and back to ε-InSe (SLs 12-15).

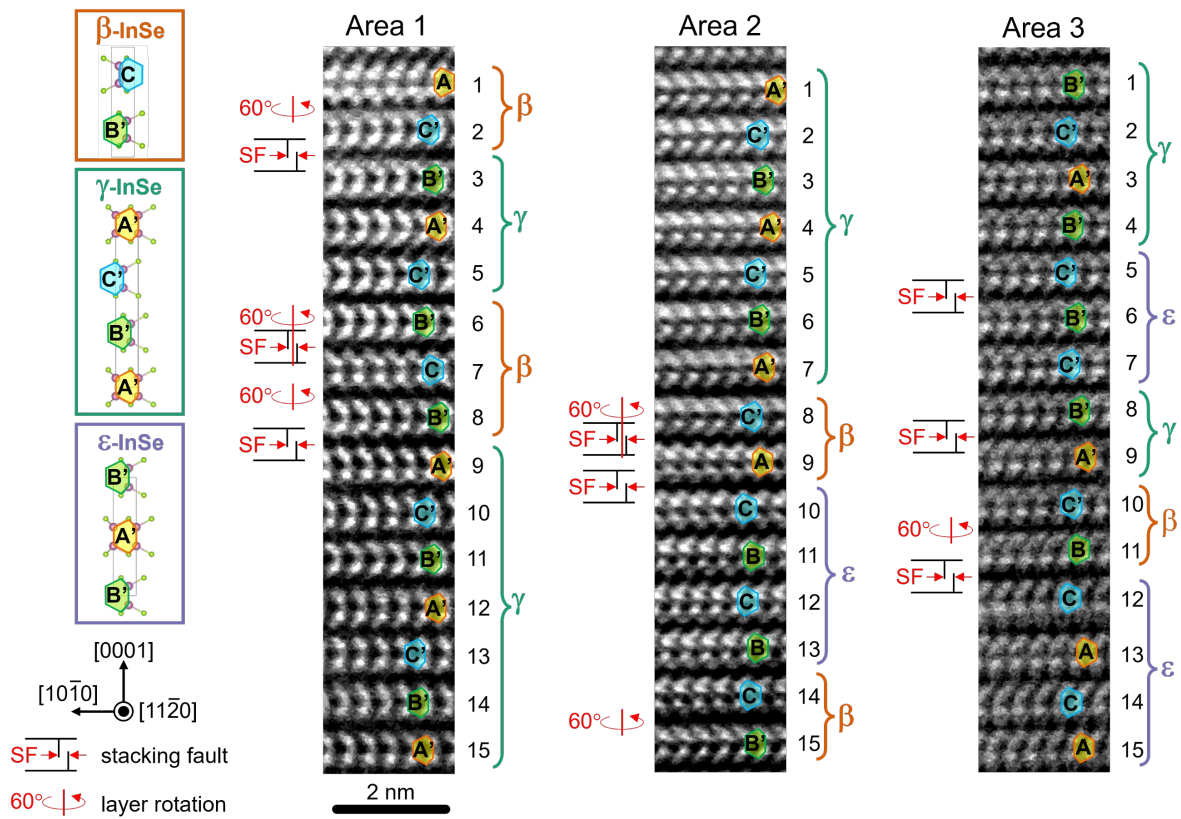

*Figure 5: Atomic scale polytype arrangement of InSe films on GaAs(111)B.* High-resolution HAADF-STEM images of three different areas in the InSe film with indicated stacking sequence A, B, and C of all visible InSe SLs and their 60°-rotational twins A', B', and C'. All three areas show stacking faults and 60°-layer rotation domains indicated by symbols on the left to each area. The resulting stacking sequence can be interpreted as domains of different polytypes in the film. A possible notation of such domains is given on the right of each area. Legends of the expected stacking sequence per each InSe polytype, zone axis and indicated defect symbols are shown on the left.

Two conclusions can be drawn from the stacking sequence analysis of the STEM images exemplarily presented in Fig. 5: (1) the different InSe polytype arrangements emerged at the nanoscale due to a high planar defect density in the film, (2) the polytype γ-InSe was found to be dominant with 52%, followed by both β-, and ε-InSe (24% for each), establishing the presence of polytype/polymorph mixtures at the nanoscale of both centrosymmetric and non-centrosymmetric phases, and rendering the non-centrosymmetric polytype γ-InSe to dominate films grown on GaAs(111)B. The nanoscale arrangement of the different polytypes further suggested that the formation energy of the different polytypes is very similar, potentially making it inherently challenging to synthesize single polytype InSe films using thin film growth approaches.

A further in-depth analysis of the HAADF-STEM images required expanding the analysis beyond the currently employed simplified picture of the known stable β-, γ- and ε-polytypes of bulk InSe.[20,25,31,47–49] Differing in their layer sequence their commonality is the alignment of the trifold In-Se bonds on either side of the In-In leading to the appearance of irregular hexagons in the [11$\bar{2}$0] projection as discussed above. In this configuration, the closest 4-atom-pair (Se-In-

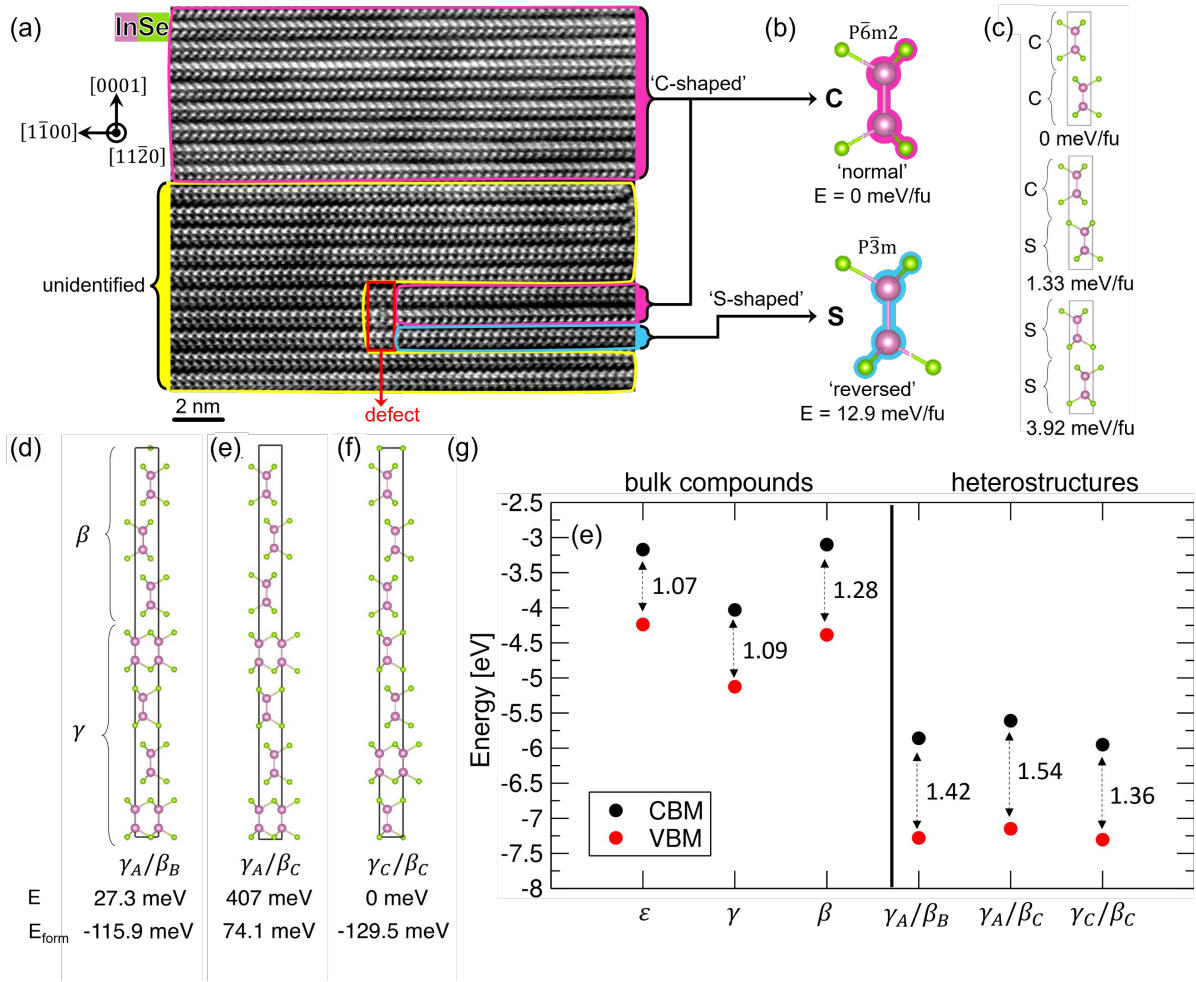

*Figure 6: Experimental observation of polymorph formation in InSe films on GaAs(111)B. (a) High-resolution HAADF-STEM image of an area in the InSe film imaged in the [**11$\bar{2}$0**] zone axis where the Se-In-In-Se bonding type in each single layer (SL) was classified as C- (pink box) or S-shaped (blue box). The bonding configuration of all SLs in the yellow box was ambiguous and was therefore classified as unidentified. A planar defect spanning across the upper and lower neighboring SLs was observed in the red box. **First principles analysis of different InSe polytypes and polymorphs.** (b) Energy difference between a C- and S-type InSe SL polymorph. (c) Energy difference and formation energies of β-InSe with 100% C-type SLs, 50% C- and 50% S-type SLs, and 100% S-type SLs. Heterostructure of γ-InSe and β-InSe interfaced at the (d) A and B, (e) A and C, and (f) C and C polyhedra. (g) Conduction band minimum (CBM) and valence band maximum (VBM) of the bulk InSe polytypes and their heterostructures. The band gap is denoted by black dotted arrow lines.*

In-Se) of each irregular hexagon chain in a SL gives rise to a 'C-shaped' contrast in HAADF-STEM. However, besides the commonly expected bonding configuration [labeled 'C-shaped' and highlighted by a pink box in Fig. 6(a)] several different bonding configurations were found in InSe films by HAADF-STEM as shown in Fig. 6(a). While not all In-Se arrangements within the layer could be unambiguously identified from the images [no clear 'C-shaped' contrast of SLs in the yellow box of Fig. 6(a)], it was found in some locations in the film that trifold In-Se bonds on one end of the In-In bond were rotated against the trifold In-Se bonds on the other end, changing the space group from the non-centrosymmetric $P\bar{6}m2$ to the centrosymmetric $P\bar{3}m$. This configuration presents as 'S-shaped' contrast in the HAADF-STEM projection and was labeled as such in the blue box in Fig. 6(a). Both C- and S-shaped bonding configurations were sketched in Fig. 6(b) for clarification. In fact, the S bonding configuration in InSe has been theoretically predicted using a swarm intelligence guided structural search.[61] Different stacking sequences of SLs with exclusively S bonding configurations were suggested, specifically the ω and φ polytypes adopting the same stacking sequence for adjacent InSe SLs like the ε- and β-polytypes, respectively. Indeed, the predicted centrosymmetric InSe layer polymorph was found in the grown InSe films, as pointed out in Fig. 6(a), however different stacking sequences from the S polymorph were not found.

Density functional theory (DFT) calculations were performed to quantify the relative formation energy differences between the different InSe SL polymorphs and polytypes experimentally found. First, the 'normal' C-shaped non-centrosymmetric quadruple InSe SL was found to be lower in energy by 12.9 meV per formula unit (fu) compared to the 'reversed' S-shaped centrosymmetric InSe SL as indicated in Fig. 6(b). When C- and S-bonded SLs were stacked as displayed in Fig. 6(c), further extremely small energy differences were obtained. In β-InSe, for example, the energy of C-C stacking was found to be the most energetically favorable [top image of Fig. 6(c)], followed by a mixed C-S stacking [1.33 meV/fu in the middle image of Fig. 6(c)], and S-S stacking being the most unfavorable [3.92 meV/fu in the bottom image of Fig. 6(c)]. These energies are much smaller than the thermal energy of about 50 meV available during the MBE film synthesis. Therefore, both layer configurations can form during non-equilibrium thin film synthesis at elevated temperature.

DFT was next used to investigate different stacking types in bulk InSe. In bulk, β-InSe was found to be the most energetically favorable, with ε- and γ-InSe higher in energy by 11.9 meV/fu and 21.7 meV/fu, respectively. Supercells containing an interface between β-InSe and γ-InSe were then constructed with different polyhedral terminations as observed experimentally in Fig. 5. As an example, the A SL of γ-InSe was placed next to either a B or C SL of β-InSe, etc.; the notation used in this case was $\gamma_X/\beta_Y$, where X and Y denote the terminating SLs of the γ- and β-phase, respectively. The three interfaces investigated in this work were $\gamma_A/\beta_B$ [Fig. 6(d)], $\gamma_A/\beta_C$ [Fig. 6(e)], and $\gamma_C/\beta_C$ [Fig. 6(f)]. Though this was not an exhaustive list of all possible interfaces, the differences in energy and electronic structure were evident.

Relative to one another, the different interfaces were not significantly different in energy. Out of those studied, the $\gamma_C/\beta_C$ structure was found to be most energetically favorable, followed by $\gamma_A/\beta_B$ [27.3 meV/fu higher in energy in Fig. 6(d) compared to Fig. 6(f)], and then $\gamma_A/\beta_B$ with a 407-meV/fu-higher energy in Fig. 6(e) compared to Fig. 6(f). Relatedly, the calculated formation energies [$E_{form}$ in Figs. 6(d), (e), and (f)] of the respective interfaces showed they can be thermodynamically favorable [-129.5 meV/fu for $\gamma_C/\beta_C$ in Fig. 6(f) and - 115.9 meV/fu for $\gamma_A/\beta_B$ in Fig. 6(d)] or unfavorable [74.1 meV/fu for $\gamma_A/\beta_C$ in Fig. 6(e)]. $\gamma_A/\beta_C$ in Fig. 6(e) likely displayed the highest energy and a positive formation energy because the Se atoms are stacked directly on top of one another across the vdW gap at this particular interface. It is therefore not surprising that MBE growth leads to a mixture of different polytypes, and polymorph microstructures given that the formation energies between them are thermodynamically favorable and the energy differences between different stacking arrangements are small.

Electronic properties of mixed polytype/polymorph InSe thin films

DFT analysis further revealed that the different binding configuration in SLs and polytype stacking arrangements exhibit different electronic band structures. In the SL, the effect was less pronounced; a band gap of 2.89 eV was calculated for the C-type non-centrosymmetric quadruple InSe SL, and a value of 2.84 eV was found for the S-type centrosymmetric InSe SL configuration. A similar small change between C- and S-type SLs was also observed in the bulk. In β-InSe with two C-type layers, for example, the band gap was calculated to be 1.28 eV. This changed to 1.20 eV upon switching of one C- to an S-SL (i.e., creating a C-S interface), and changed further to 1.09 eV when both layers were S-type, resulting in a 0.2-eV-difference compared to β-InSe with only C-type SLs.

As shown in Fig. 6(g), bulk ε-, γ-, and β-InSe were computed to have band gaps of 1.07 eV, 1.09 eV, and 1.28 eV, respectively, in relative agreement with previous experimental and computational results.[62,63] Interestingly, while the band edges [valence band maximum (VBM) and conduction band minimum (CBM) in Fig. 6(g)] were similar between the ε and β phases (CBM of -3.17 eV/-3.10 eV and VBM of -4.24 eV/-4.38 eV for ε-/ β -InSe, respectively), there are large band offsets between those two phases and the γ-polytypes (CBM of -4.03 eV and VBM of -5.12 eV), as is shown in Fig. 6(g). There is therefore a band alignment between ε- InSe and β-InSe, but a type 2 band offset between ε- or β-InSe and γ-InSe, leading to possible electronic inhomogeneity. Furthermore, when β-InSe and γ-InSe were interfaced, the stacking sequence also influenced the electronic structure. The band gaps in Fig. 6(g) of the interfaces $\gamma_A/\beta_B$, $\gamma_A/\beta_C$, and $\gamma_C/\beta_C$ [in analogy to Figs. 6(d), (e), and (f)] were larger than either constituent component by itself, ranging from 1.36 eV to 1.54 eV depending on the relative stacking between the two phases. The band edges also shift to significantly more negative CBM/VBM values of -5.86 eV/-7.27 eV for $\gamma_A/\beta_B$, -5.61/-7.14 eV $\gamma_A/\beta_C$, and -5.95 eV/-7.30 eV $\gamma_C/\beta_C$. Therefore, even within the heterostructures of different stackings, there are different electronic structures.

These DFT calculations revealed that throughout film growth, it is energetically feasible to form InSe with both differently bonded SLs (C- and S-type) and different stacking sequences. Consequently, this dramatically affects the electronic structure, which will be locally different throughout the entire sample. The nanoscale polytype arrangement not only gave rise to a mixed polar/non-polar domain arrangement, the different polytypes separated by sizeable band offsets are furthermore expected to induce electronic disorder. The nanoscale energy barrier structure caused by the electronic disorder likely forms spatially separated electron and hole pockets in the film that may dominate the overall electronic transport properties.

Two representative four-point probe sheet resistance measurements performed on 40-nm-thick InSe thin film Hall bar devices A and B as sketched in Fig. 7(a) are shown in Fig. 7(b). At low temperatures the film resistance exceeded the sheet resistance values that can be measured.

Only at room temperature (RT) and above sheet resistance values between $10^9$ Ω/sq. (300K, which was found comparable to the resistance of the substrate at RT by placing four electrical contacts on the exposed substrate around the Hall bar structure[64]) to mid-$10^5$ Ω/sq. were detected. The temperature dependence of the resistance for both devices indicated that the films were highly insulating, suggesting that the density of itinerant carriers in the film was very low. Estimating the intrinsic carrier concentration $n_i$ of the γ-InSe polytype from the effective density of states $N_{DOS} = 2 \cdot (2\pi\, m_{eff}\, k_B T\, h^{-2})^{\frac{3}{2}}$ and using for $m_{eff}$ the electron and hole effective masses for γ-InSe $m_e = 0.14 \cdot m_0$ and $m_h = 2.3 \cdot m_0$ with $m_0$ being the free electron mass, $k_B$ and $h$ Boltzmann and Planck's constant, we find for the conduction and valence band density of states $N_{CB} = 2.5 \cdot 10^{14} \cdot T^{\frac{3}{2}}$ cm$^{-3}$ and $N_{VB} = 1.7 \cdot 10^{16} \cdot T^{\frac{3}{2}}$ cm$^{-3}$. The intrinsic carrier concentration $n_i = \sqrt{N_{CB} \cdot N_{VB}} \cdot \text{Exp}[-E_g/(2k_BT)]$ at 300K and 400K thus gives $7.5 \cdot 10^9$ cm$^{-3}$ and $2.2 \cdot 10^{12}$ cm$^{-3}$, certainly approaching the lower detection limit of the electrical measurement setup at RT. However, a typical unintentional carrier concentration on the order of $10^{17}$ cm$^{-3}$ and electron mobilities of around 1000 cm²/Vs at RT resulting in sheet resistance values on the order of tens of k$\Omega$/sq were expected even at RT as such values were reported for InSe grown by pulsed laser deposition, and field effect transistors made from exfoliated and chemical vapor transport synthesized and encapsulated InSe.[19,28,29] The much higher observed sheet resistance thus pointed towards two possible interpretations; a) the introduction of a much lower content of unintentional carriers during the MBE growth process compared to other reported methods, and b) the presence of electronic disorder in the InSe film introduced by the mixed polytype/polymorph nanoscale domains that dominated the transport characteristics by rendering the expected much higher number of unintentional free carriers immobile, as well as a combination of a) and b). It seemed that at higher temperature a hopping-like transport [the resistivity was found to scale linearly with $T^{-1/3}$ in Fig. 7(b) corroborating interpretation b) assuming a transport process similar to the hopping regime observed in lightly doped semiconductors but at much enlarged scales[65]] was enabled allowing sufficient carriers to be thermally excited and overcoming the energy barrier separating the different polytypes. Hall voltages $V_H$ measured on the two voltage lead pairs H1 and H2 of the Hall bar device scheme depicted in Fig. 7(a) for devices A and B at 400 K are shown in Figs. 7(c) and 7(d), respectively. From the direction of the magnetic field and sign of Hall voltage changing linearly with applied magnetic field the carriers in device A were determined to be electrons, consistent with literature reports.[24,32,47] Note the good agreement of the Hall voltage curves taken at the Hall contact pair H1 and H2 for device A in Fig. 7(c). A film carrier concentration of $3.7 \cdot 10^{15}$ cm$^{-3}$ and carrier mobility of about 700 cm$^2$/Vs was determined at 400 K, in good agreement with reports for Bridgman-synthesized bulk $\gamma$-InSe.[66] While the high carrier mobility at 400 K underlined the possible potential of InSe for low-power, high-performance electronics, the at least three-orders-of-magnitude-higher than intrinsic observed carrier concentration confirmed that a significant amount of unintentional carriers as introduced into the InSe film either during the MBE growth process itself or during the Hall bar device fabrication. For device B and in contrast to device A, however, a strong deviation from the linear Hall effect was observed. Furthermore, the data taken at the Hall contact pairs H1 and H2 looked different, suggesting a much larger degree of electronic disorder. The change in slope of the Hall voltage was indicative of electron and hole carriers contributing to transport. The electron-like behavior dominating the Hall effect at low magnetic fields and a hole-like trend at higher magnetic fields suggested a higher carrier mobility for electrons than for holes. While the observed transport properties were found to confirm the presence of unintentional carriers in MBE-grown InSe, as well as the presence of electronic disorder introduced by the nanoscale polytype/polymorph mixture within the films as predicted by DFT, i.e., interpretations a) and b), the origin of the unintentional carriers remained concealed.

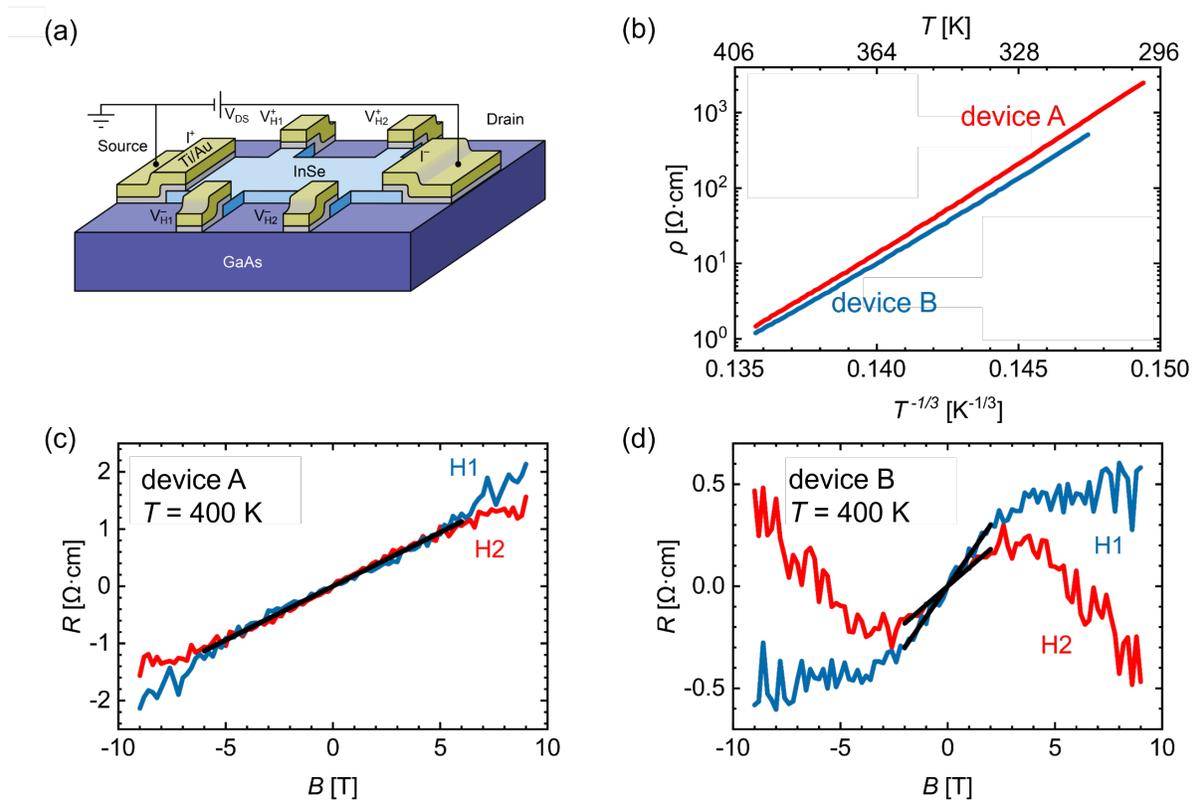

*Figure 7: Effect of nanoscale InSe polytype/polymorph domains on electronic transport properties. (a) Schematic of the Hall bar device structure used to measure electronic transport properties of InSe thin films on GaAs(111)B. (b) Temperature-dependence of the InSe film resistance measured for two Hall bar devices A and B from 300K to 400K. Hall coefficient versus magnetic field measured in (c) device A and (d) device B at 400K across the two Hall voltage contact pairs H1 and H2 in (a).*

**Conclusion**

In summary, InSe films grown on GaAs(111)B by MBE formed different polytypes and polymorphs assuming polar, non-centrosymmetric as well non-polar, centrosymmetric arrangements within the individual SLs in InSe and through different layer stacking sequences. This nanoscale polytype domain structure was accompanied by an inherent energetic disorder. The formation of the different polytypes suggested that their formation energy is very similar, which was confirmed by DFT calculations. The electronic band structure alterations across the different polytypes and polymorphs suggested an energetic disorder in InSe films that lead to a suppression of transport at room temperature and signified that energy barriers emerging at polytype/polymorph domain boundaries dominated the electronic transport characteristics in these films. Our combined experimental and theoretical results unveil potential challenges associated with a bottom-up synthesis approach to grow single polytype InSe films using thin film techniques operating far from equilibrium conditions, but also hold promise to realize InSe polytypes and polymorphs that are energetically less favorable.

**Methods**

InSe thin film synthesis

Thin InSe films were grown on undoped, semi-insulating GaAs(111)B substrates purchased epi-ready from AXT. Substrates were loaded into a R450 MBE reactor from DCA Instruments with a base pressure of $4\times10^{-10}$ Torr. Native oxide removal was obtained by heating the as-loaded GaAs(111)B wafers to 400 °C while exposed for 60 minutes to a reactive hydrogen flux supplied from a HABS[67,68] source (Karl Eberl MBE Komponenten), operated at a filament heater current of 14 A and a hydrogen background pressure of $9.3\times10^{-7}$ Torr. InSe films were grown at a sample temperature of 350 °C. In and Se fluxes were generated by conventional effusion cells and were measured by a quartz crystal microbalance (QCM) from Colnatec. In and Se fluxes of $4.7\times10^{13}$ $cm^{-2}s^{-1}$ and $1.2\times10^{14}$ $cm^{-2}s^{-1}$ were used, respectively. Associated tooling factors for QCM flux measurements for In and Se were obtained by physical film thickness measurements using X-ray diffraction (XRD) on 10 – 50 nm thick pure Se and In films. At this growth temperature the close to three times higher Se flux was necessary to compensate for the loss of the more volatile Se from the film's growth front. Film growth was performed for 30 minutes. Reflection high energy electron diffraction (RHEED) images were taken during native oxide removal and throughout the growth.

X-ray diffraction

XRD was carried out ex-situ with a Panalytical X'Pert[3] four-circle diffractometer in high resolution configuration using a PIXcel 3D detector and CuK$_{\alpha1}$ radiation. The XRD optics consisted of a hybrid Ge(220) crystal monochromator with a 1/32 ° slit and a 10 mm mask, clipping the X-ray beam to a 20-mm-long line illuminating the entire width of the (10×10) mm samples with a thickness of 1.83 mm to 0.13 mm for small and large diffraction angles, respectively.

## Atomic force microscopy

The film morphology was analyzed ex-situ with a Dimension Icon Bruker atomic microscope by mapping the surface in Peak-Force Tapping mode using Scanasyst-air tips in the ScanAsyst in air instrument configuration.

## Raman spectroscopy

Raman spectroscopy measurements were performed using a Horiba LabRam system with unpolarized 488 nm laser excitation (7.1 mW of total power) with a neutral power density filter of 25 % in backscattering geometry. The laser was focused through a 100× objective in backscattering geometry and cut by an additional notch filter to ±10 cm$^{-1}$ using a spectral resolution set by the grating of 1800 g mm$^{-1}$.

The Raman mapping was performed on the same Raman setup using an excitation laser wavelength of 532 nm operated at 34 mW power and a neutral power density filter of 1% focused through a 50× objective lens.

## Transmission electron microscopy

High-resolution scanning transmission electron microscopy (STEM) were taken at 300 kV in cross section using a dual spherical aberration-corrected FEI Titan$^3$ G2 60-300 S/TEM. All STEM images were recorded with the beam propagating along the $[\bar{1}10]$ zone axis of GaAs –i.e. the $[11\bar{2}0]$ azimuth of InSe – using a high-angle annular dark field (HAADF) detector with a collection angle of 50-100 mrad. Energy-dispersive X-ray spectroscopy (EDS) maps were collected using the Super-X, four quadrant SDD EDS system and Bruker Espirit software on the Titan microscope. Cross-sectional TEM specimen were prepared using an FEI Helios 660 focused ion beam (FIB) system. A thick protective amorphous carbon layer was deposited over the region of interest. A beam of Ga+ ions was used in the FIB to make the specimen electron transparent for TEM images. Initially a kinetic energy of 30 keV was used for the Ga beam, which was then stepped down to 1 kV to avoid ion beam damage to the specimen surface.

Density functional theory

Spin-polarized density functional theory (DFT) calculations using the Vienna ab initio Simulation Packag (VASP)[69,70] with the projector augmented wave (PAW) pseudopotentials.[71,72] The "strongly constrained and appropriately normed" (SCAN) meta-generalized-gradient approximation (meta-GGA) was used, with van der Waals interactions included using the rVV10 vdW density functional (i.e., SCAN+rvv10).[73] In the InSe monolayers, a minimum 15 Å of vacuum was included to separate the periodic images. All calculations utilized a 900-eV plane wave cutoff. The following Monkhorst-Pack[74] k-point meshes were used: 12x12x1 were for the monolayers, 8x8x2 for the bulk compounds, and 14x14x2 for the stacked heterostructures. During optimizations, the lattice parameters and ionic positions were relaxed until the forces on the atoms were smaller than 0.01 eV/Å. To correct the underestimation of the band gap, the Heyd–Scuseria–Ernzerhof (HSE) exchange–correlation functional was used,[75] with 35% exact exchange included. This amount was determined by computing the band gap of $\beta$-InSe with varying amounts of exact exchange until agreement with the experimental band gap was met.

Electronic transport property measurements

To measure the electrical properties of InSe films grown on GaAs(111)B, Hall bars were fabricated by photolithography using a tri-layer resist stack of polymethylmethacrylate (PMMA), polydimethylglutarimide (PMGI), and another positive spin-coating photoresist positive resist to avoid exposing the InSe film to any alkali developers, which can react with a transition metal chalcogenide.[76,77] The film was etched with reactive ion etching (RIE) to remove the unprotected film using a mixture of Ar, $Cl_2$ and $CF_4$ before the protective resist stack was removed, resulting in a Hall-bar device of (55×20) μm in size with 40 μm between the centers of the voltage leads (H1 and H2), which in turn were 5 μm in width. The same photolithography process was used to define electrodes by depositing 5-nm-thick Ti film followed by a 45-nm-thick Au cap, schematically shown in Fig. 7(a). All transport devices shown here were prepared on the same InSe film deposited on a GaAs wafer. Different devices were measured in a Physical Properties Measurement System (PPMS) equipped with an 8 T superconducting magnet in the temperature range from 4K to 400K. The DC source bias was provided by a Keithley 6340 Sub-Femtoampere Remote Source Meter, four-point and Hall voltages were measured using two Keithley 2182A nanovoltmeters.

## Data availability

All data contained in this work is available during the review process under the following link …, This private link will be converted to an open-access link to ScholarSphere with stand-alone DOI for data supporting this work upon publication.

## Acknowledgements

Film growth, characterization, device, and manuscript preparation of this work were supported by the National Science Foundation through the Penn State 2D Crystal Consortium-Materials

Innovation Platform (2DCC-MIP) under NSF cooperative agreements DMR-1539916, and DMR-2039351. Supporting experimental work for substrate de-oxidation, film characterization, electrical measurements and data analysis was supported by the US Department of Energy, Office of Science, Office of Basic Energy Sciences Energy Frontier Research Centers program under Award Number DE-SC0021118. M.L. and J.Y. were supported by startup funds provided by the Newark College of Engineering at the New Jersey Institute of Technology. DFT calculations were performed on the Wulver cluster at the New Jersey Institute of Technology and on the Carbon cluster at the Center for Nanoscale Materials, Argonne National Laboratory. Work performed at the Center for Nanoscale Materials, a U.S. Department of Energy Office of Science User Facility, was supported by the U.S. DOE, Office of Basic Energy Sciences, under Contract No. DE-AC02-06CH11357. The Carbon allocation CNM 83203 was used for this work. We gratefully acknowledge useful discussion with Prof. Tom Jackson.

## Author contributions

M.H. and R.E.-H. conceived and designed the growth experiments. M.H. synthesized the films with help from D.S.H.L., performed XRD and AFM measurements. M.H. performed Raman spectroscopy on the film, J.G. performed HAADF-STEM experiments. J.R. and S.D. fabricated the devices and performed the electrical measurements on the devices. Jinyuan Y. measured Raman on the devices and analyzed the Raman and electrical transport data obtained from the devices. M.L. performed DFT calculations under the supervision of J.Y.. M.H., J. R., J.Y., Y.L. and R.E-H. performed the data analysis and prepared the manuscript. All authors provided their input and approved the final version of the manuscript.

## Competing interests

Authors declare no competing interests.